\documentclass[aps,pra, twocolumn,superscriptaddress,twocolumn]{revtex4-1} 
\usepackage{amsmath}
\usepackage{latexsym}
\usepackage{graphicx}
\usepackage{amsfonts}
\usepackage{braket}
\usepackage{natbib}
\usepackage{amssymb}
\usepackage{fancyhdr}
\usepackage[utf8]{inputenc}
\usepackage{dsfont}
\usepackage{colortbl}
\usepackage{xcolor}
\usepackage{subfigure}
\usepackage{psfrag}
\usepackage[colorlinks=true,citecolor=blue,linkcolor=blue]{hyperref}
\usepackage{tikz}
\usepackage{ulem}
\usetikzlibrary{arrows,shapes}

\newcommand{\LR}[1]{\left(#1\right)}

\definecolor{DarkGreen}{rgb}{0.0,0.5,0.2}

\begin{document}

\title{Engineering adiabaticity at an avoided crossing with optimal control}
\author{T. Chasseur}
\author{L. S. Theis}
\affiliation{Universit\"at des Saarlandes, Saarbr\"ucken, Germany}
\author{Y. R. Sanders}
\affiliation{IQC and Dept. of Physics and Astronomy, University of Waterloo, 200 University Ave. W, Waterloo, ON, N2L 3G1, Canada}
\author{D. J. Egger}
\affiliation{Universit\"at des Saarlandes, Saarbr\"ucken, Germany}
\author{F. K. Wilhelm}
\affiliation{Universit\"at des Saarlandes, Saarbr\"ucken, Germany}
\affiliation{IQC and Dept. of Physics and Astronomy, University of Waterloo, 200 University Ave. W, Waterloo, ON, N2L 3G1, Canada}

\begin{abstract}
We investigate ways to optimize adiabaticity and diabaticity in the Landau-Zener model with non-uniform sweeps. We show how diabaticity can be engineered with a pulse consisting of a linear  sweep augmented by an oscillating term. We show that the oscillation leads to jumps in populations whose value can be accurately modeled using a  model of multiple, photon-assisted Landau-Zener transitions, which generalizes work by Wubs \emph{et al.} [New J. Phys. \textbf{7}, 218 (2005)]. We extend the study on diabaticity using methods derived from optimal control. We also show how to preserve adiabaticity with optimal pulses at limited time, finding a non-uniform quantum speed limit.
\end{abstract}

\maketitle

\section{Introduction}
The adiabatic theorem, which should be applied with care \cite{Marzlin_PRL_93_160408, Sarandy_QIP_3_331, Tong_PLA_339_288}, states that if the time evolution of a quantum system is sufficiently slow, transitions between eigenstates can be neglected. It is thus a statement about an approximation rather than a rigorous theorem in the mathematical sense. Adiabatic quantum computing is one paradigmatic example of the usefulness of such time evolutions \cite{Farhi_arXiv,Childs_PRA_65_012322}. Another application concerns the control of quantum processor elements by frequency tuning. On the one hand, gate designs often rely on adiabaticity \cite{DiCarlo09,Martinis_PRA_90_022307}. On the other hand, due to spurious couplings --- e.g. higher-order interactions beyond nearest neighbours to other parts of the chip \cite{Galiautdinov_PRA_85_042321} and undesired spurious resonators \cite{Simmonds_PRL_93_077003, Whittaker_PRB_90_024513, Cole_APL_97_252501, Sousa_PRB_80_094515} --- and adiabatic following (i.e., reach a nonadiabatic sweep within the limited bandwidth of a realistic experiment), one desires a detour to diabaticity. Shortcuts to adiabaticity, i.e. arriving at the same final state as the adiabatic evolution but in a shorter time, have been investigated \cite{Torrontegui_AAMOP_62_117}. The study of adiabaticity in periodically driven systems has led to adiabatic Floquet theory \cite{Breuer_PLA_140_507, Guerin_PRA_55_1262, Guerin_PRA_56_1458}. The physics of adiabaticity is well captured in the Landau-Zener (LZ) model which analytically describes the behavior of a system when linearly swept through an avoided level crossing \cite{Landau_PZS_2_46, Zener_PRSL_137_696}. The generic nature of this model gives it a wide range of applications, for instance in transitionless quantum driving of spins \cite{Berry_JPAMT_42_365303}. Considering more complex, i.e. non-linear, pulses has useful applications, such as Landau-Zener-St\"uckelberg interferometry. This allows one to determine the parameters of the avoided crossing by quickly sweeping back and forth through it \cite{Berns_Nature_445_51, Shevchenko_492_1, Forster_PRL_112_116803}. The LZ model also describes tunneling states in the tunneling model of amorphous solids. Here an atom can move between two adjacent positions separated by a potential barrier. Therefore, non adiabatic driving of tunneling states affects the dielectric constant in glasses \cite{Ludwig_PRL_90_105501, Nalbach_JLTP_137_395}. Evidence for interacting defects in glasses has been presented in \cite{Carruzzo_PRB_50_6685}. Studies of the magnetisation of molecular magnets \cite{Bogani_NatMat_7_179} involve a similar situation where a magnetic field is swept over many spins \cite{Chiorescu_JMMM_221_103}. These dynamics also occur in molecular collisions \cite{Nakamura02}.

Parallel to these developments are those in quantum optimal control where a control pulse is shaped to realize a specific time evolution \cite{Rice_Book, Brumer_Book}. These methods were originally pioneered for nuclear magnetic resonance and have started to gain popularity in solid state quantum information devices. For instance in electron spin qubits to engineer gates \cite{Cerfontaine_arXiv}, as well as in superconducting qubits \cite{Clark_Nature_453_1031, Devoret_Science_339_1169} to address leakage \cite{Motzoi_PRL_103_110501, Lucero_PRA_82_042339}, frequency crowding \cite{Schutjens_PRA_88_052330, Vesterinen_arXiv} and gate design \cite{Egger_SUST_27_014001, Kelly_in_prep}. In such systems the crucial entangling gates are realized with an anti-crossing \cite{Barends_Nature_508_500}. Within the framework of qubits and optimal control, the LZ model has been used to study quantum speed limits \cite{Schulman_LNP_734_107, Vaidman_AJP_60_182, Caneva_PRL_103_240501} and how robust high fidelity pulses are to uncertainties in the non-controllable part of the Hamiltonian \cite{Grace_PRA_85_052313}.

For the detour to diabaticity we investigate the dynamics of a LZ system under a linear sweep augmented by a fast oscillation. A similar system has been studied in \cite{Wubs_NJP_7_218, Zhong_PRA_90_023635}. We investigate how such a pulse can be used to engineer diabaticity. Optimal control methods allow us to deepen this study as well as investigate  
pulses that keep the evolution adiabatic \cite{Torrontegui_AAMOP_62_117} which are crucial to quantum computing. The work is structured in the following way. Section \ref{Sec:Diabat} discusses the LZ system when the linear sweep is augmented by an oscillating term. The analytics are in Sec. \ref{Sec:Diabat_An} whilst the numerics are presented in Sec. \ref{Sec:Diabat_Num}. The adiabatic pulses are discussed in Sec. \ref{Sec:adiabat}.

\section{Engineering Diabaticity \label{Sec:Diabat}}

The LZ Hamiltonian is defined by $\hat H_\text{LZ}=\varepsilon(t)\hat\sigma_z/2+\Delta\hat \sigma_x/2$ where $\hat\sigma_x$ and $\hat \sigma_y$ are Pauli matrices in the basis states $\ket{0}$ and $\ket{1}$, which also are eigenstates in the absence of the coupling $\Delta$. A finite $\Delta$ mixes these states into new energy eigenstates referred to as instantaneous eigenstates and introduces an avoided level crossing localized at $\varepsilon(t)=0$. 

Sweeping through that anti-crossing with constant velocity $v$  following $\varepsilon(t)=vt$ results in the well celebrated transition probability between $\ket{0}$ at $t\rightarrow -\infty$ and $\ket{1}$ and $t\rightarrow\infty$

\begin{equation} \label{Eqn:LZ_Prob}
 P_\text{LZ}=1-\exp\left(-\frac{\pi\Delta^2}{2v}\right)\,.
\end{equation}
Therefore an linear sweep at high speed, $v\gg\Delta^2$ , avoids leakage in our basis of $\ket{0}$ and $\ket{1}$ hence keeping the time evolution {\em diabatic}, whereas an infinitely slow sweep, $v\ll\Delta^2$, keeps the system in the instantaneous ground (or excited) state at all times while that state is changing.

\subsection{Analytics of an Oscillation-Augmented Linear Sweep\label{Sec:Diabat_An}}
Consider a linear sweep augmented by an oscillation of amplitude $\lambda$ and frequency $\Omega$
\begin{equation} \label{Eqn:OscillationPulse}
 \varepsilon(t)=vt+\lambda\cos\LR{\Omega t +\varphi}\,.
\end{equation}
The phase $\varphi$ determines the value of the oscillation when the linear sweep is on resonance with the anti-crossing and plays an important role as emphasized below. This form of $\varepsilon$ can arise when investigating the dielectric constant of glasses by using an external electric field \cite{Ludwig_PRL_90_105501, Nalbach_JLTP_137_395}. 
 Wubs \emph{et al.} \cite{Wubs_NJP_7_218} study a very similar model where $\varepsilon$ is linear only and $\Delta$ varies sinusoidally in time. We will see that similar physics arises in our model and will connect to their work when appropriate. To study the time evolution produced by Eq. (\ref{Eqn:OscillationPulse}) we switch to the interaction picture by the transformation $\hat U_0=\exp\{-i\phi(t)\hat\sigma_z/2\}$ resulting in
\begin{align} \notag
 \hat H_\text{I}=\frac{\Delta}{2}\begin{pmatrix} 0 & e^{i\phi(t)} \\ e^{-i\phi(t)} & 0\end{pmatrix}~~~\text{with}~~~\phi(t)=\int_0^{\,t}{\rm d}\tau\,\varepsilon(\tau)\,.
\end{align}
Using the Taylor expansion of $\sin$ and $\exp$ and the binomial expansion of $(e^{i\Omega t}-e^{-i\Omega t})^n$ these terms are cast into the form
\begin{align} \label{Eqn:Sum_expand}
 e^{\pm i\phi(t)}=\sum\limits_{m=-\infty}^\infty(\pm)^mJ_m\LR{\frac{\lambda}{\Omega}}e^{\pm i\frac{1}{2}vt^2+im(\Omega t+\varphi)}\,.
\end{align}
$J_m$ is the $m^\text{th}$ Bessel function of first kind. This expansion is similar to that done in \cite{Shevchenko_492_1, Goorden_PRB_68_0125508} and is motivated by coherent destruction of tunneling \cite{Grossmann_PRL_67_516, Grifoni_PR_304_229, Wubs_CP_375_163}. The expansion in Eq. (\ref{Eqn:Sum_expand}) allows one to identify term $m$ with an $|m|$-photon emission/absorption process \cite{Ashhab_PRA_75_063414}. Using the property $J_{-m}(x)=(-1)^mJ_m(x)$ the Hamiltonian in the interaction picture is
\begin{align} \label{Eqn:H_int}
 \hat H_\text{I}=&~\frac{\Delta}{2}\sum\limits_{m=-\infty}^\infty J_m\LR{\frac{\lambda}{\Omega}}\times \\ \notag
 &~\left[\cos\LR{m\varphi}\begin{pmatrix}0 & e^{\frac{i}{2}vt^2+im\Omega t} \\ e^{-\frac{i}{2}vt^2-im\Omega t} & 0 \end{pmatrix}\right. \\ \notag
 -&~\left.i\sin\LR{m\varphi}\begin{pmatrix}0 & -e^{\frac{i}{2}vt^2+im\Omega t} \\ e^{-\frac{i}{2}vt^2-im\Omega t} & 0 \end{pmatrix}\right]\,.
\end{align}
For $\lambda=0$ the dynamics under a linear sweep are recovered. Under these conditions, the system undergoes a transition when $t=0$ and the final population is given by Eq. (\ref{Eqn:LZ_Prob}). This jump is a result of the function $vt^2$ having an extrema at $t=0$ and around this time the Hamiltonian is no longer rapidly oscillating. Thus if a Magnus expansion were done \cite{Warren_JCP_81_5437}, one would see that the higher order commutators cannot be neglected; these contributions add up producing a jump. A Dyson series expansion also fails since as $\varepsilon(t)$ can be zero, there is no small parameter to expand in.

\begin{figure}[htbp!] \centering
 \includegraphics[width=0.45\textwidth]{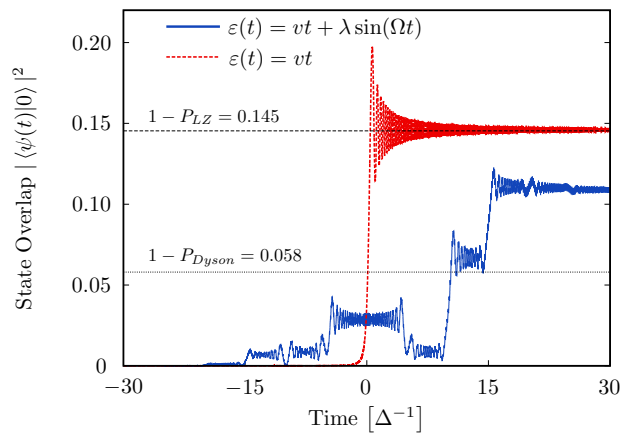}
 \caption{Comparison of a linear sweep with one augmented by an oscillation. The time evolution was computed from $-T/2$ to $T/2$ with $T=100\Delta^{-1}$ thus cutting off times without significant population changes. The linear speed was $v=10\Delta^{2}$, the oscillation frequency and amplitude were $\Omega=50\Delta$ and $\lambda=120.24\Delta$ respectively. The population jumps occur at $t=-m\Omega/v$ which here are integer multiples of $5\Delta^{-1}$. The dotted line is the transition probability that would be expected of a first order Dyson series expansion. As expected, it does not predict the correct probability as the expansion is not valid since $\varepsilon$ can be smaller than $\Delta$. \label{Fig:Diabat_Ex}}
\end{figure}

Another way to see this is to note that Eq. (\ref{Eqn:H_int}) can be viewed as a sum of Landau-Zener Hamiltonians taken in the interaction picture where the zero bias point is shifted from $\varepsilon=0$ to $\varepsilon=m\Omega$. Indeed the terms $\pm i(vt^2/2+m\Omega t)$ can be recast into the form
\begin{align} \label{Eqn:Jump_Times}
 \pm i\frac{v}{2}\LR{t+\frac{m\Omega}{v}}^2\mp i\frac{m^2\Omega^2}{2v}\,.
\end{align}
We thus expect many jumps to happen at intervals of $m\Omega/v$ which is illustrated in Fig. \ref{Fig:Diabat_Ex}. This generalizes the results of Ref.
\cite{Wubs_NJP_7_218} where only two separate jumps appear. 

\subsubsection{Multi-Jump Model}
\begin{figure}[htbp!] \centering
 \includegraphics[width=0.45\textwidth]{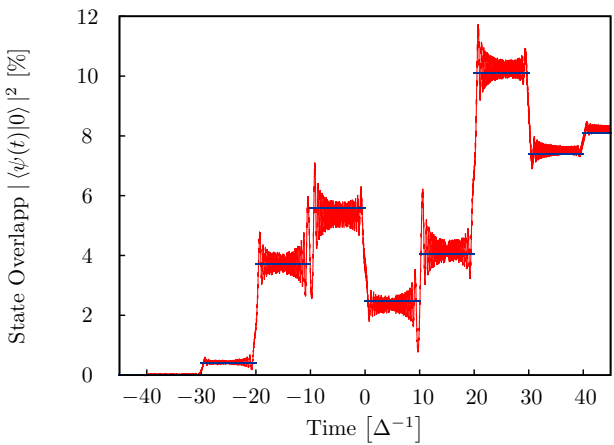}
 \caption{Illustration of the multi-jump model. The red line shows the full time propagation. The horizontal lines show the state overlapp as predicted by the multi-jump model $\hat U_\text{\tiny J}$. As can be seen this model accurately predicts the photon assisted LZ transitions. \label{Fig:multi_jump}}
\end{figure}

Building on the observations of the previous section we construct an approximate model that allows to 
analytically compute the unitary matrix describing the jumps at $m\Omega/v$. It consists of using the separation into multiple Landau-Zener Hamiltonians and assume their application is non\textcolor{red}{-}overlapping in time, i.e., that the jumps are independent and that the sweep between the jumps only contributes a phase factor, which assumes $\Delta\ll\Omega$. A time evolution that spans up to $m_0$ photon jumps is thus approximated by
\begin{align} \label{Eqn:UniJump}
 \hat U_\text{\tiny J}=\prod\limits_{m=-m_0}^{m_0}\begin{pmatrix} e^{-\frac{\pi\Delta_m^2}{4v}} & e^{i\tilde\varphi_m}\sqrt{1-e^{-\frac{\pi\Delta_m^2}{2v}}} \\ e^{-i\tilde\varphi_m}\sqrt{1-e^{-\frac{\pi\Delta_m^2}{2v}}} & e^{-\frac{\pi\Delta_m^2}{4v}} \end{pmatrix}\,.
\end{align}
The terms in the matrices are similar to those from the LZ transition probability in Eq. (\ref{Eqn:LZ_Prob}) but jump $m$ has $\Delta_m=\Delta J_\text{m}(\lambda/\Omega)$ instead of $\Delta$ and an associated Stokes phase \cite{Wubs_NJP_7_218, Shevchenko_492_1} given by
\begin{align} \notag
 \tilde\varphi_m\simeq-\frac{\Delta^2}{4v}\ln\LR{\frac{T^2v}{4}}-\frac{\pi}{4}-\frac{m^2\Omega^2}{2v}-m\varphi
\end{align}
  under the assumption $\Delta^2/4v\leq1$. Note that negative $\Delta_m$ occur, leading to  an additional phase factor of $\pi$. This model properly accounts for the jump heights as shown in Fig. \ref{Fig:multi_jump} but is completely devoid of the oscillations around these values as is to be expected given that each LZ jump matrix in $\hat U_\text{\tiny J}$ is time independent.

\subsubsection{Engineering Diabaticity}

Now the challenge of engineering diabaticity is to sweep through the avoided level crossing such that at the start and end of the evolution we stay in the same uncoupled state, i.e. the time evolution operator up to an irrelevant global phase should be $\hat U(-T/2,T/2)=\mathds{1}$. In Ref. [\onlinecite{Wubs_NJP_7_218}] it is shown how time-reversal anti-symmetry can lead to a supression of population transfer yielding $\hat U=\mathds{1}$. This can be done if the time evolution for $t>0$ reverses the time evolution of $t<0$ which can be related to the Loschmidt echo \cite{Jalabert_PRL_86_2490}. In this section we connect this idea  to our model and show how one has to carefully set the ratio $\lambda/\Omega$ to a particular value.

\begin{figure}[htbp!] \centering
 \includegraphics[width=0.45\textwidth]{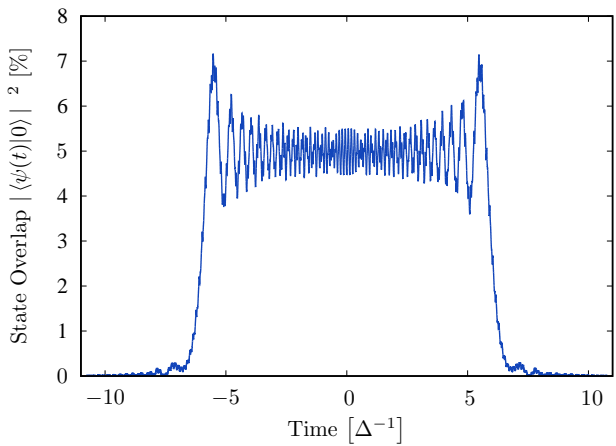}
 \caption{Example of a pulse with $v=8\Delta^2$, $\Omega=50\Delta$ and a time dependent $\lambda(t)$ given by Eq. (\ref{Eqn:Lambda_of_t}). The ramp and switching times are given by $T+T_\text{s}=3\Omega/v+T_\text{s}=21.55\Delta^{-1}$. The ramp height satisfies $\lambda_\text{r}=2.4048\Omega$ so as to cancel the jump at $m=0$. The final occupation of $\ket{0}$ is smaller than $10^{-5}$. \label{Fig:Diabat_Engin}}
\end{figure}

\begin{figure}[htbp!] \centering
 \includegraphics[width=0.45\textwidth]{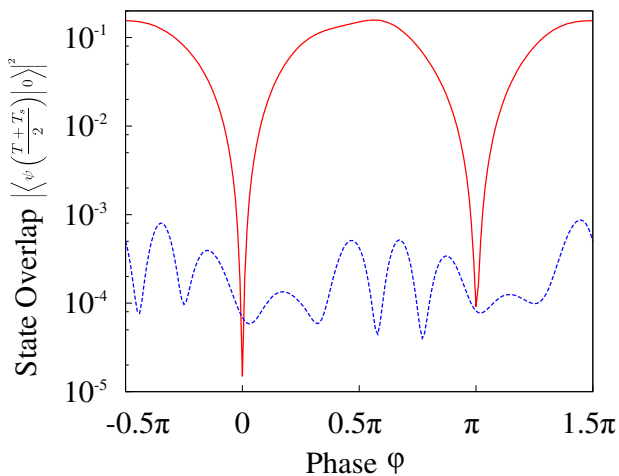}
 \caption{The red solid line shows phase's effect on the state overlap at $T/2$ of the pulse in Fig. \ref{Fig:Diabat_Engin}. A wrongly tuned phase induces large errors. The dashed blue line corresponds to a shorter pulse where the oscillation is switched off before the $m=\pm1$ jumps can occure. As expected, the error is much less sensitive to the phase. \label{Fig:Diabat_Engin_Phase}}
\end{figure}

This can be achieved with high accuracy if the time during which the oscillation is present is chosen so that only the photon assisted jumps at $m=\pm1$ occur. One has to set the phase $\varphi$ to zero and choose the ratio $\lambda/\Omega$ to the first zero of the zeroth Bessel function $J_0$, i.e. $\lambda/\Omega\simeq 2.4048$. This suppresses the transition at $t=0$ and the photon assisted jumps at $m=\pm1$ cancel eachother. This follows from the Bessel function property $J_{-1}(x)=-J_1(x)$. To include the finite rise time of any electronics, we choose a linear ramp for the oscillation amplitude
\begin{align} \label{Eqn:Lambda_of_t}
 \lambda(t)=\left\{\begin{array}{l l}
                    \lambda_\text{r} & \text{if}~|t|< \frac{T-T_\text{s}}{2} \\[0.5em]
                    \frac{\lambda_\text{r}}{T_\text{s}} (\frac{T+T_\text{s}}{2}-|t|) & \text{if}~\frac{T-T_\text{s}}{2} \leq |t|\leq \frac{T+T_\text{s}}{2} \\[0.5em]
                    0 & \text{otherwise}
                   \end{array}
\right.
\end{align}
$T_\text{s}$ is the switching time. The height of the ramp is chosen to suppress the jump at $m=0$, thus $J_0(\lambda_\text{r}/\Omega)=0$ is needed implying $\lambda_\text{r}\cong2.4048\,\Omega$. The duration $T-T_s$ over which the amplitude of the oscillation is held constant has to be long enough to include the $m=\pm1$ transitions but the ramp duration plus switching time $T+T_\text{s}$ has to be chosen so that there are no longer any oscillations in the pulse for $|t|>2\Omega/v$ so as to prevent $|m|\geq2$ jumps. This imposes $(T-T_s)/2>\Omega/v$ and $(T+T_\text{s})/2<2\Omega/v$. The time evolution of such a pulse is shown in Fig. \ref{Fig:Diabat_Engin}. There is no jump at $t=0$ and the $m=\pm 1$ jumps cancel each other. As expected, this results in a pulse which leaves hardly any population in $\ket{0}$ namely less than $10^{-5}$. However this is phase sensitive.Indeed if $\varphi\neq0$ then  
the Stokes phases of $m=\pm 1$ are no longer different by $\pi$ and the reasoning presented above no longer holds. This is shown in Fig. \ref{Fig:Diabat_Engin_Phase} by the solid red line. At its worst, the phase can produce almost 16\% error. This can be understood that only at phase $0$, the tunneling events assisted by an odd number of electrons vanish automatically due to destructive interference of time-reversed paths. 

If even shorter pulses can be made, the phase variable can be rendered irrelevant. Indeed, for pulses shorter than $\Omega/v$, only the population jump at $m=0$ could contribute as no photon-assisted processes are resonant. However, it is removed by the right choice of $\lambda/\Omega$. The error of a shorter pulse with $T+T_\text{s}=13.86\Delta^{-1}$ is shown in Fig. \ref{Fig:Diabat_Engin_Phase} by the dashed blue line.

\subsection{Numerical optimal control \label{Sec:Diabat_Num}}
The previous section showed that a linear sweep with an oscillation can reduce population jumps when going through the anti-crossing compared to a simple linear sweep. Here we show how a numerical optimization of the four parameters of the pulse in Eq. (\ref{Eqn:OscillationPulse}) can reduce an unwanted transition when $T>2\Omega/v$ and without windowing. Next we show that starting from a linear sweep a time-sliced gradient search provides an optimal solution with a smaller oscillation but with a chirped frequency.

\subsubsection{Optimization of a Linear Sweep with a Single Frequency Oscillation Pulse}
\begin{figure}[htbp!] \centering
 \includegraphics[width=0.48\textwidth]{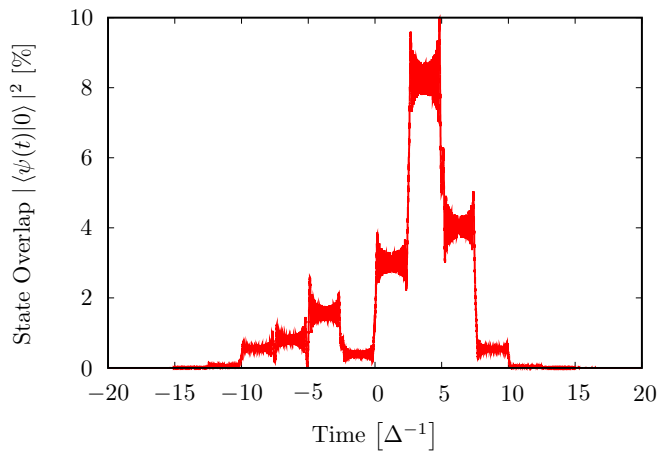}
 \caption{Time evolution under the action of a pulse given by Eq. (\ref{Eqn:OscillationPulse}) where the parameters were optimized with the NM algorithm. The pulse duration was set to $T=200\,\Delta^{-1}$ and the optimization resulted in $\Omega=99.9718\,\Delta$, $\lambda=311.631\,\Delta$, $v=9.99409\Delta^2$ and $\varphi=2\pi\,0.387$. 
At the end of the pulse, the error is of order $10^{-6}$. \label{Fig:NM_Pulse}}
\end{figure}

When the short window of Eq. (\ref{Eqn:Lambda_of_t}) cannot be created experimentally, diabaticity can still be engineered by optimizing the four parameters $(v,\lambda,\Omega,\varphi)$ of Eq. (\ref{Eqn:OscillationPulse}). This is done with the Nelder-Mead (NM) simplex search algorithm \cite{Nelder_CompJ_7_308} which is often used in optimal control \cite{Biercuk_Nature_458_996, Doria_PRL_106_190501, Kelly_in_prep, Egger_PRL_112_240503}. The metric to be minimized is $|\braket{\psi(T)|0}|^2$. Pulses of arbitrarily low population transfer from $\ket{1}$ to $\ket{0}$ can be found with NM. Figure \ref{Fig:NM_Pulse} shows a pulse with performance perfect down to $10^{-6}$ where the optimization was stopped -- in principle, machine precision can be reached. This pulse does not rely on suppressing the $m=0$ transition or time-reversal anti-symmetry. Instead, the parameters found allow many photon assisted transitions but they produce a time evolution that leaves no population in $\ket{0}$ at the end of the pulse by tayloring their interference properly. Numerically we found good convergence regardless of the initial parameters, but typically our analytical solutions from the previous sections were used as initial guesses. Many high quality pulses having different parameters can be found.

\begin{figure}[htbp!] \centering
 \includegraphics[width=0.44\textwidth]{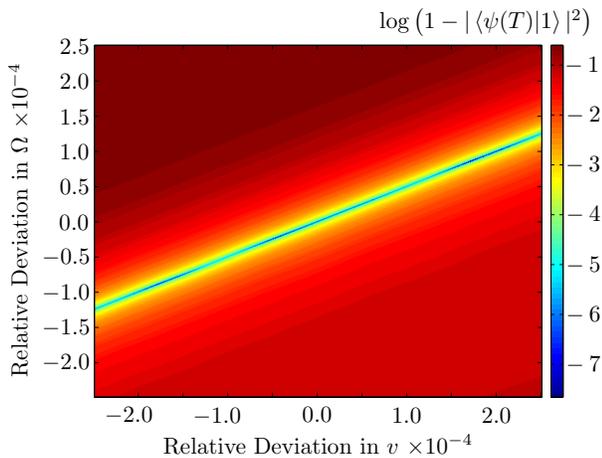}
 \caption{Error in logarithmic scale as function of the error in the parameters $(v,\Omega)$ relative to the optimum found by the NM algorithm for the pulse in Fig. \ref{Fig:NM_Pulse}. As can be seen their is a narrow valley where errors on $v$ and $\Omega$ do not affect the fiedelity too severly. However this valley is very steep signifying that the parameters $\Omega^2$ and $v$ need to have the right ratio. \label{Fig:FullPulse_Robust}}
\end{figure}

In an experiment the parameters of the pulse can differ from those intended. It is thus important to characterize how robust the pulse is with respect to parameter fluctuations. To study how robust these pulses are we introduce small errors on the optimal values and study the decrease in fidelity. We find that these pulses are very sensitive to small errors in the parameters $v$ and $\Omega$ as shown in Fig. \ref{Fig:FullPulse_Robust}. The pulse is only sensitive to errors in the ratio between $\Omega^2$ and $v$, not to these parameters separately.  This can be traced back to the observation that the Stokes phase in Eq. (\ref{Eqn:UniJump}) strongly depends on $\Omega^2/v$. For parameters $\varphi$ and $\lambda$ we find that the optimal pulse is less sensitive and can tolerate roughly an order of magnitude more fluctuations than in $v$ and $\Omega$. Such errors in parameter values could be corrected by using the experiment to close the control loop \cite{Egger_PRL_112_240503}.

\subsubsection{GRAPE pulses}
To go beyond the constraint imposed by the pulse shape of Eq. (\ref{Eqn:OscillationPulse}) we use the optimal control algorithm named Gradient Ascent Pulse Shape Engineering (GRAPE). It tries to optimize a fidelity $\Phi$ which is a functional of the pulse $\varepsilon(t)$. The pulse sequence is discretized into $N$ constant pixels $u_j$ of time $\Delta T=N/T$. GRAPE proceeds by iteratively updating the pixel values according to the rule $u^{(n)}_j\mapsto u_j^{(n+1)}=u_j^{(n)}+\epsilon\nabla_j\Phi$ where $\nabla_j\Phi$ is the gradient of the fidelity function with respect to pixel $j$. Details of the procedure are given in \cite{Khaneja_JMR_172_296305}. The gradient is computed analytically \cite{Machnes_PRA_84_022305}. For our situation we use the gate overlap fidelity
\begin{align} \label{Eqn:Phi}
 \Phi=\frac{1}{4}\left\vert\text{Tr}\left(\hat\sigma_0^\dagger\hat U[\varepsilon(t)]\right)\right\vert^2
\end{align}
where $\sigma_0$ is the identity matrix and $\hat U[\varepsilon(t)]$ is the time evolution realized by the pulse $\varepsilon(t)$.

\begin{figure} 
 \includegraphics[width=0.48\textwidth]{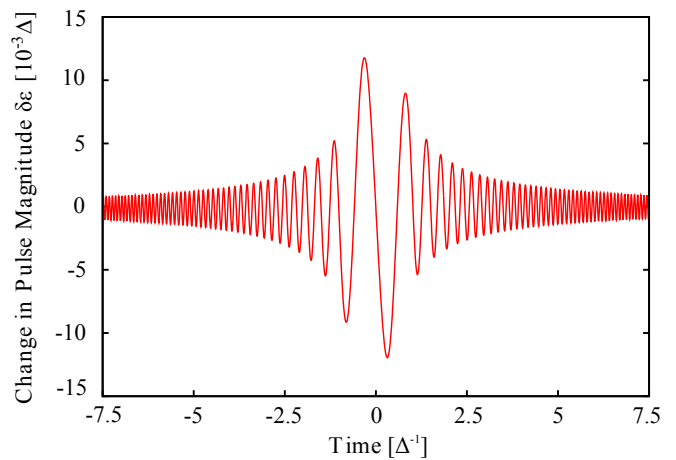}
 \caption{Pulse change due to GRAPE on an initial linear sweep of duration $9.6\Delta^{-1}$ and $v=40\pi^2\Delta^2$. The figure shows the central region of the pulse where the change is most pronounced. Time and pulse amplitude are respectively given in inverse and proportional units of the gap size $\Delta$. \label{Fig:GRAPE_Pulse}}
\end{figure}

When considering the LZ problem we use a linear sweep with speed $v$ without coherent drive as initial pulse. GRAPE achieved the target error of $10^{-5}$ by adding to the initial pulse a modulation with a time dependent frequency. This extra modulation is shown in Fig. \ref{Fig:GRAPE_Pulse}. Its amplitude is much smaller than the fixed frequency case. The new oscillation is reminiscent of the St\"uckelberg oscillations with decaying amplitude and increasing frequency. This can be viewed as a rotating frame version of the oscillation-augmented sweep - instead of sweeping the energy splitting, the frequency is swept accordingly. 

\section{Quantum Speed Limits for an adiabatic evolution \label{Sec:adiabat}}
This section considers the case when we wish to transfer the population between the two bare states by sweeping through the anti-crossing. Equivalently this corresponds to staying in the same energy branch at all times. This can be achieved if the time evolution is adiabatic. Time evolutions of this sort are important in quantum computing where one wishes to exchange a quanta between two quantum elements such as a resonator and a qubit or two qubits \cite{Mariantoni_Sci_334_6165}. Recently a protocol has been demonstrated that allows fast adiabatic two-qubit gates by making use of only $\sigma_z$ control and optimal window functions \cite{Martinis_PRA_90_022307}. For quantum computing, the population occupation and the phases can be important.

\subsection{Adiabatic Pulses}
\begin{figure}[htbp]
  \includegraphics[width=0.48\textwidth]{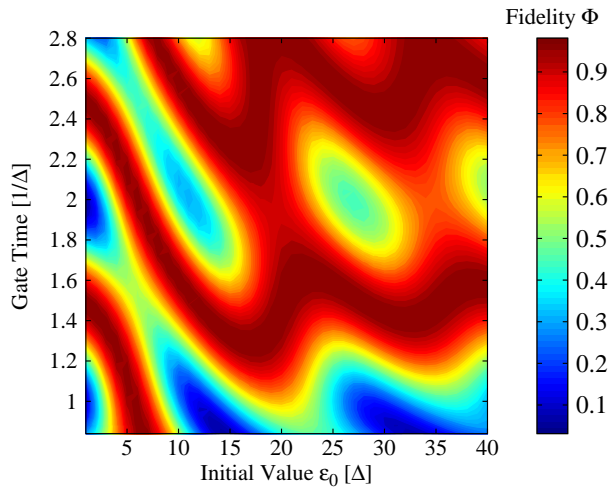}
 \caption{Local phase insensitive fidelity for the initial pulse given by equation (\ref{Eqn:start_pulse}) for different gate times and start values for $\Delta=0.04\;{\rm GHz}$  in all simulations. This shows that this start pulse fails to keep the population in the energy eigenstate whilst crossing the anticrossing, i.e. it is not adiabatic. \label{Fig:QSL_Scan_Init}}
\end{figure}

In this section we wish to realize the gate $\hat U_\text{des.}=\hat\sigma_x$ through a pulse with boundary conditions
$\varepsilon(\pm T/2)=\pm\varepsilon_0$. The intial pulse for the GRAPE optimization is
\begin{align} \label{Eqn:start_pulse}
 \varepsilon(t)=-\Delta\tan\left[\frac{\arctan\LR{\frac{\Delta}{\varepsilon_0}}-\frac{\pi}{2}}{{\rm erf}\LR{-\lambda\frac{T}{2}}}{\rm erf}\LR{\lambda t}\right]
\end{align}
The parameter $\lambda T$ controls the width of the sweep hence $\lambda$ alone sets the speed at which the point $\varepsilon=0$ is crossed. The pulse presented in Eq. (\ref{Eqn:start_pulse}) is an already studied pulse form designed so that the sweep across the anti-crossing is done slowely, but further away from the anticrossing the sweep velocity increases \cite{Martinis_PRA_90_022307}. Figure \ref{Fig:QSL_Scan_Init} shows the phase insensitive fidelity, defined by $\Phi_\text{ins}=(|\braket{0|\hat U|1}|^2+|\braket{1|\hat U|0}|^2)/2$ , of the initial pulse in Eq. (\ref{Eqn:start_pulse}) as a function of $(\varepsilon_0,T)$. $\Phi_\text{ins}$ only measures the population transfer. There are regions where the pulse does not perform as well as expected even when the gate time is long.

\subsection{Numerics}
\begin{figure}[htbp!] \centering
 \includegraphics[width=0.48\textwidth]{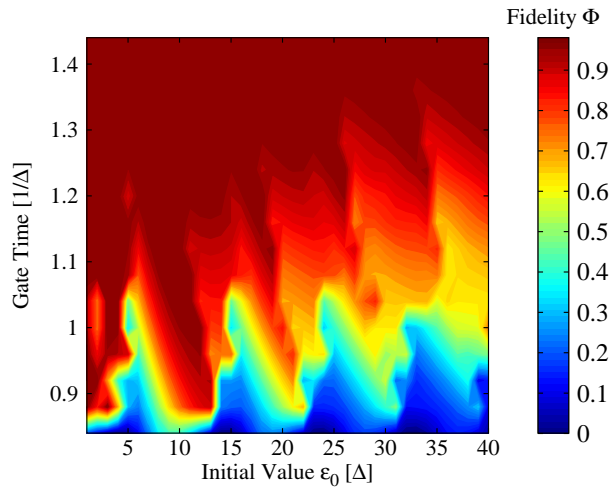}
 \caption{Fidelity of the pulses optimized with GRAPE as a function of duration and sweep range for 
 $\Delta=0.04\;{\rm GHz}$. Once the gate time is sufficently large, any fidelity can be reached. 
 \label{Fig:QSL_Scan}}
 \end{figure}

\begin{figure}[htbp!] \centering
 \includegraphics[width=0.48\textwidth]{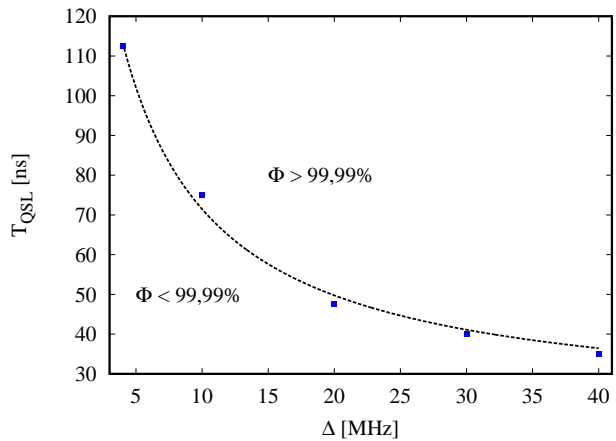}
 \caption{Study of the quantum speed limit after optimization of the pulse with the GRAPE algorithm. For each specific $\Delta$ a gate time is found for which  a high fidelity can be achieved for almost all values of $\varepsilon_0$. The data points in blue are then fitted by the dashed line corresponding to Eq. (\ref{Eqn:Fit}).\label{Fig:QSL}}
\end{figure}

To improve on the situation shown in Fig. \ref{Fig:QSL_Scan_Init} we use the GRAPE algorithm and study the dependence of the fidelity on gate time $T$. To preserve the adiabatic nature of the processs, the pulse is convoluted with a Gaussian to remove any fast oscillations. The fidelity to be optimized is the same as in Eq. (\ref{Eqn:Phi}) but with the target gate being $\hat\sigma_x$ instead of the identity. This guarantees that both phase and population are correct after the gate. Fig. \ref{Fig:QSL_Scan} depicts the final gate fidelity $\Phi$ in the $(\varepsilon_0,T)$ parameter space where each pulse corresponds to the result of a GRAPE optimization starting from Eq. (\ref{Eqn:start_pulse}). As can be seen all the regions of Fig. \ref{Fig:QSL_Scan_Init} that showed a loss of fidelity have been improved on. This is true as long as the pulse is long enough. Below a certain duration a form of quantum speed limit (QSL) is encountered and GRAPE can no longer improve the fidelity. The dependence of this QSL is further studied in Fig. \ref{Fig:QSL}. It shows the behavior of the QSL as function of $\Delta$. The figure was created by choosing a gate time for which a good fidelity ($>99.99\%$) can be found for almost all values of $\mathcal{\varepsilon}_0$. The data points (blue squares) are then fitted to
\begin{align} \label{Eqn:Fit}
 T_\text{QSL}(\Delta)=t_0+\frac{c}{\Delta+\Delta_0}
\end{align}
with $c$ and $\Delta_0$ as fit parameters whereas $t_0$ is controlled by the fact that we have used "buffer pixels" in GRAPE \cite{Egger_SUST_27_014001} to prevent steep initial rise and final drop. As seen from the figure this empirical fit describes very well the behaviour observed for $T_\text{QSL}$.

\pagebreak

\section{Conclusion}
We have analytically discussed the influence of a linear sweep augmented by fast and strong single frequency oscillation on the dynamics of a two level system. We showed how such pulses can be used to engineer diabaticity without resorting to large bandwidth control. Optimal control can help go beyond the analytic considerations producing pulses with machine precision level error. In the adiabatic study, improved adiabatic pulses were found with optimal control and their quantum speed limit was discussed.

\section{Acknowledgments}
We wish to thank J. Martinis and E. Sete for discussions on adiabatic pulses and C. Koch for a great conversation on adiabatic pulses. This work was supported by the EU through SCALEQIT and QUAINT
and funded by the Office of the Director of National Intelligence (ODNI), Intelligence Advanced Research Projects Activity (IARPA), through the Army Research Office. All statements of fact, opinion
or conclusions contained herein are those of the authors and should not be
construed as representing the official views or policies of IARPA, the ODNI,
or the U.S. Government.

\bibliography{EngineeringAdiabaticity.bib}
\bibliographystyle{apsrev4-1}

\end{document}